# Organic chemistry in a $CO_2$ rich early Earth atmosphere

Benjamin Fleury[1*], Nathalie Carrasco[1,2], Maëva Millan[1], Ludovic Vettier[1], Cyril Szopa[1,2]

[1] Université Versailles St-Quentin ; Sorbonne Universités, UPMC Univ. Paris 06 ; CNRS/INSU, LATMOS-IPSL, 11 Boulevard d'Alembert, 78280 Guyancourt, France

[2] Institut Universitaire de France, 103 Bvd St-Michel, 75005 Paris, France

[*]Corresponding author. Current address:

Benjamin Fleury

Jet Propulsion Laboratory, California Institute of Technology

4800 Oak Grove Drive, MS 183-301, Pasadena, CA 91109, USA

Mail: benjamin.fleury@jpl.nasa.gov

Pages: 30

Tables: 3

Figures: 7

# Abstract

The emergence of life on the Earth has required a prior organic chemistry leading to the formation of prebiotic molecules. The origin and the evolution of the organic matter on the early Earth is not yet firmly understood. Several hypotheses, possibly complementary, are considered. They can be divided in two categories: endogenous and exogenous sources. In this work we investigate the contribution of a specific endogenous source: the organic chemistry occurring in the ionosphere of the early Earth where the significant VUV contribution of the young Sun involved an efficient formation of reactive species. We address the issue whether this chemistry can lead to the formation of complex organic compounds with $CO_2$ as only source of carbon in an early atmosphere made of $N_2$, $CO_2$ and $H_2$, by mimicking experimentally this type of chemistry using a low pressure plasma reactor. By analyzing the gaseous phase composition, we strictly identified the formation of $H_2O$, $NH_3$, $N_2O$ and $C_2N_2$. The formation of a solid organic phase is also observed, confirming the possibility to trigger organic chemistry in the upper atmosphere of the early Earth. The identification of Nitrogen-bearing chemical functions in the solid highlights the possibility for an efficient ionospheric chemistry to provide prebiotic material on the early Earth.

Key words: Atmospheres, chemistry; Organic chemistry; Prebiotic chemistry; Earth

# 1. Introduction

Life is supposed to have appeared on Earth before 3.5 Ga during the Archean and maybe before the Late Heavy Bombardment (LHB) during the Hadean (Nisbet and Sleep, 2001). The presence of liquid water and organic matter is now widely accepted as conditions for the apparition of life (Cottin et al., 2015). Understanding the origins of these organic molecules, which are implicated in the apparition of life is of prime interest for astrobiology. Different origins are proposed for the organic matter on the early Earth at this period: exogenous delivery by meteorites and comets, or endogenous formation in hydrothermal vents or in the primitive atmosphere (Chyba and Sagan, 1992).

Reduced atmospheres are known to be an important source of organic matter as pointed out by the observation of Titan the largest satellite of Saturn (Tomasko and West, 2010; Waite et al., 2010) or by experiments realized by Miller about the reactivity of the early Earth atmosphere (Miller, 1953). However, if the question of the degree of oxidation of the early Earth atmosphere and so its composition is not completely solved, there is evidence that the upper mantle of the Earth was at the present redox state since 3.9 Ga (Delano, 2001) and probably 4.4 Ga (Trail, 2011) resulting in a relatively oxidant primitive atmosphere dominated by molecular nitrogen $N_2$ and carbon dioxide $CO_2$ (Kasting, 1993). The composition of the early Earth's atmosphere is an important parameter to consider for the production of organic matter. Different experiments highlighting the fact that chemistry in oxidant atmospheres produces less organic compounds than in reduced atmospheres (Schlesinger and Miller, 1983a, b).

Recent experimental works showed that organic aerosols can be produced in an atmosphere dominated by $CO_2$ with a minor amount of methane $CH_4$ (Trainer et al., 2004; Trainer et al., 2006). But it remains uncertain whether $CH_4$ was present in the primitive atmosphere of the Earth. Indeed, today 70% of the methane emissions have biogenic sources and for an important part from ecosystems, which do not exist during the Hadean and the Archean eons (Denman et al., 2007). The two principal abiotic sources of methane considered for the early Earth are delivery by meteoritic and cometary impacts and production by serpentinization process (Kasting, 2005). These sources could be responsible for a methane amount of only few $ppm_v$ in the primitive atmosphere of the Earth (Emmanuel and Ague, 2007; Feulner, 2012; Guzmán-Marmolejo et al., 2013; Kasting, 2005), which is much lower than the amount considered in experimental simulations (Trainer et al., 2006) or climate model (Charnay et al., 2013). For these reasons, we have chosen to explore a potential organic chemistry in an early Earth's atmosphere with only $CO_2$ as source of carbon.

Moreover, if $CH_4$ concentration should be low in the early Earth's atmosphere, molecular hydrogen $H_2$ could be present in a larger amount. Indeed, assuming that the loss of hydrogen to space was limited by diffusion, a mixing ratio of $10^{-3}$ has been established for the early Earth's atmosphere (Kasting, 1993). But, assuming a lower Jeans escape than today, Tian, 2005 has proposed that $H_2$ mixing ratio could be 100 times higher, reaching 10% and up to 30%. According to the authors, this lower efficiency of the Jeans escape for $H_2$ would be explained by a lower temperature of the exobase at this period. Indeed, in one hand the higher $CO_2$ concentration at this period would result to a higher radiations backed to space and in the other hand, the lower oxygen concentration in the atmosphere ($O_2$) would result to a lower heating of the exobase by the atomic oxygen UV absorbance (Martin et al., 2006; Tian et al., 2005). This new calculation has been discussed since, notably because of the limited amount of $CO_2$ present in the atmosphere to cool the exobase (Catling, 2006; Tian et al., 2006). A more recent model has determined a hydrogen mixing ratio of 1% at the homopause. It results that the exact concentration of $H_2$ during the Hadean and the Archean remains difficult to determine but could be important. For this reason, the influence of $H_2$ on the primitive atmospheric reactivity needs to be studied.

We present in this paper an experimental study of the organic growth occurring in an early Earth's atmosphere made of $N_2$, $CO_2$ and $H_2$. A first study performed in similar conditions highlighted an important formation of water vapor in the stratosphere and ionosphere of the early Earth (Fleury et al., 2015). We are now interested in investigating the concomitant formation of organic molecules, as both gaseous and solid products. The composition of the Earth atmosphere has evolved over geological time, notably $CO_2$ whose concentration during the Hadean and the Archean is not precisely known (Feulner, 2012). For this reason, we also take into account the influence of the $CO_2$ initial amount on the atmospheric chemistry of the early Earth simulated in our experiments.

## 2. Experimental setup and analytical protocols

### 2.1. Experimental simulation

We used here the PAMPRE experimental setup, which has been described in details in previous publication (Szopa et al., 2006). It is a Radio Frequency Capacitively Coupled Plasma (RF CCP) at low pressure. In this experiment, a discharge is generated between a polarized electrode and a cylindrical grid grounded electrode confining the plasma. Before each experiment, the reactor is heated and pumped down to $2 \times 10^{-6}$ mbar. The gaseous flow is then injected continuously and pumped through a rotary vane vacuum pump. Three gas bottles are used in the experiments to generate the reactive gaseous mixture: one with high-purity of $N_2$ (99.999%), one containing a $N_2$-$H_2$ mixture with 5% of $H_2$ and one with $CO_2$ (> 99.995%).

The generator RF power is set to 30 W and the total gas flow rate to 55 sccm resulting in a 0.9 mbar pressure in the reactor. The modeling of the used plasma has been done in pure $N_2$ (Alves et al., 2012). The electrons energy distribution function (EEDF) presents a maximum at 2 eV and a relatively populated tail for electron energy above 4 eV and up to 14 eV mimicking the young solar spectrum, which presents a higher UV and X-ray emission flux and a lower visible and infrared emission flux than today (Claire et al., 2012). The electrons present in our plasma have sufficient energy to dissociate and ionize $N_2$ and activate nitrogen chemistry as well as $CO_2$ and $H_2$. The PAMPRE experiment can be so used to simulate the reactivity of the upper layers (above the troposphere) of the early Earth, where $N_2$, $CO_2$ and $H_2$ can be dissociated and ionized.

The hydrogen mixing ratio is kept constant at 4% for all experiments. This concentration is chosen in agreement with recent modeling studies of the $H_2$ mixing ratio, giving an upper limit about 1% for the Archean atmosphere (Kuramoto et al., 2013) and up to 30% (Tian et al., 2005). The gas flow is adjusted among $N_2$ and $CO_2$ from an experiment to another to introduce $CO_2$ at different mixing ratios: 1%, 5% and 10%.

## 2.2. Cryogenic trapping

In order to detect and identify gas species produced in low quantity, we have trapped the volatile species and accumulated them by cooling the plasma box. For that, a part of the experiments presented below is performed with a continuous injection of liquid nitrogen ($L_{N2}$) inside the stainless-steel block supporting the grounded electrode. The plasma box is cooled by thermal conduction. The temperature is fixed at 173 K to prevent $N_2$, $H_2$ and $CO_2$ condensation. However, the trapping of $CO_2$ has been observed in the results presented in the section 3.4. The trapping of the reactants is discussed in the same section and has been take into account for the interpretation of the results. Products are accumulated during 4 or 8 hours of plasma duration. After plasma turning off, the reactor is pumped to eliminate $N_2$, $H_2$ and $CO_2$. The reactor is isolated and the plasma box is warmed up to room temperature. Trapped gases are released and analyzed by mass spectrometry, infrared spectroscopy and by Gas Chromatography coupled to Mass Spectrometry (GC-MS).

### 2.3. Mass spectrometry on gaseous phase

Measurements of the gas phase trapped during a part of the experiments, as described in section 2.1, are achieved with a Pfeiffer QME 200 quadrupole mass spectrometer (QMS). In the spectrometer, neutral species are ionized by electron impact at 70 eV. Gases are transferred to the spectrometer through a capillary tube, which is long enough to keep the pressure below $10^{-5}$ mbar in the spectrometer when the pressure inside the reactor is 0.9 mbar. Its resolution is 100 at *m/z* 100 and it covers 1-100 u mass range. Gas trapped are analyzed with a long scanning acquisition of 1 s for each mass between *m/z* 1 and 60.

### 2.4. Infrared spectroscopy on gaseous phase

Trapped molecules are also analyzed with a Thermo Scientific Nicolet 6700 Fourier Transform Infrared (FTIR) spectrometer. A schema of the FTIR setup on the PAMPRE reactor, seen from above, is presented in Figure 1. The infrared beam is emitted by the FTIR source and passes through the reactor via two KBr windows. Then, the beam is collected by a Mercury Cadmium Telluride (MCT) detector cooled by liquid nitrogen.

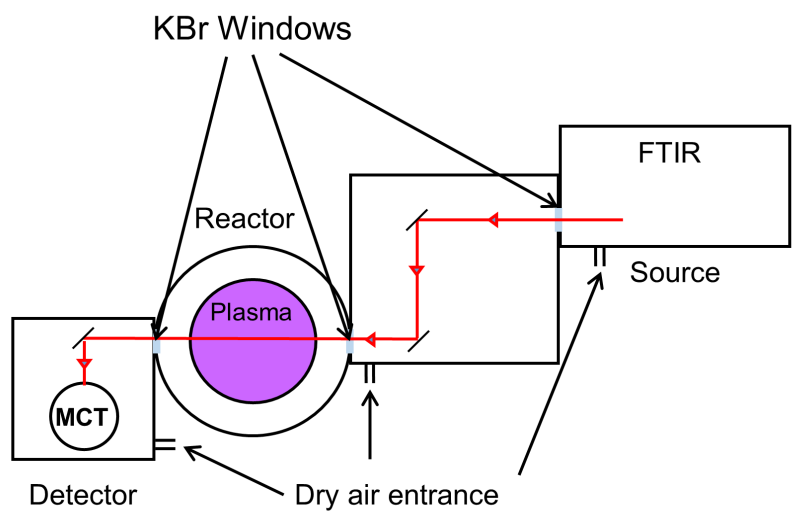
Figure 1: Schema of the FTIR setup on the PAMPRE reactor.

In the results presented below, IR spectra are recorded in the 650-4500 cm$^{-1}$ range with a resolution of 1 cm$^{-1}$ after a co-addition of 500 scans. With only one passage of the beam through the reactor, the corresponding path length is 50.8 ± 0.2 cm.

### 2.5. GC-MS analysis of the gaseous phase

To analyze the gaseous phase *ex situ* by GC-MS, we transfer the trapped gases in the reactor into another external cold trap. It is a cylindrical glass coil immersed in liquid nitrogen and connected to the reactor. Before transferring gases, the cold trap is pumped down to $5 \times 10^{-5}$ mbar. Then, the valve isolating the trap from the reactor is opened and the gaseous species are transferred into the external cold trap. No formation of a solid residue is observed through the warming process here, contrarily to an experiment made in Titan-like conditions (Gautier et al., 2011).

The GC-MS analyses of the gas trapped are achieved using a Thermo Scientific trace GC ultra with an ITQ Thermo Scientific mass spectrometer. Gases are injected through a six port gas valve. The mass spectrometer is composed by a quadrupole using a 70 eV electron ionization system. For the gas separation, the column is a MXT-QPLOT (Restek, 30m long, 0.25 mm internal diameter and 10 μm stationary phase thickness). The column temperature is set at 40°C during 5 min, then the temperature is increased with a gradient of 5 °C/min up to 190°C and kept at this final temperature for 5 min. Helium is used as the carrier gas (>99.9995% purity) at a constant 1.5 ml/min flow rate. A blank is done before each sample analysis.

## 2.6. Solid phase collection and infrared analysis of thin films

$CaF_2$ substrates are placed on the grounded electrode to collect thin films possibly produced during the experiments. These potential films are deposited at room temperature during 40 hours of plasma duration. Then, they are analyzed by infrared spectroscopy (Attenuated Total Reflectance technique). Samples are placed on the surface of a prism with a high refraction index (ATR crystal). The infrared signal is collected by a Deuterium TriGlycine Sulfate (DTGS) detector in the 1200-4000 $cm^{-1}$ range with a resolution of 4 $cm^{-1}$ after a co-addition of 500 scans.

## 3. Results

### 3.1. $CO_2$ and $H_2$ consumption

The first aspect of the gas phase reactivity involves the consumption of the three initial species: $N_2$, $CO_2$ and $H_2$. The $N_2$ consumption is low, with a dissociation of about 4%, theoretically evaluated in (Alves et al., 2012). The consumptions of $H_2$ and $CO_2$ are more important and are monitored at room temperature by *in situ* mass spectrometry using the time-tracking of $CO_2^+$ at *m/z* 44 and $H_2^+$ at *m/z* 2 at a time resolution of 0.5 s. The time-evolution of $H_2$ is given on Figure 2, and the time-evolution of $CO_2$ has been previously studied (Fleury et al., 2015).

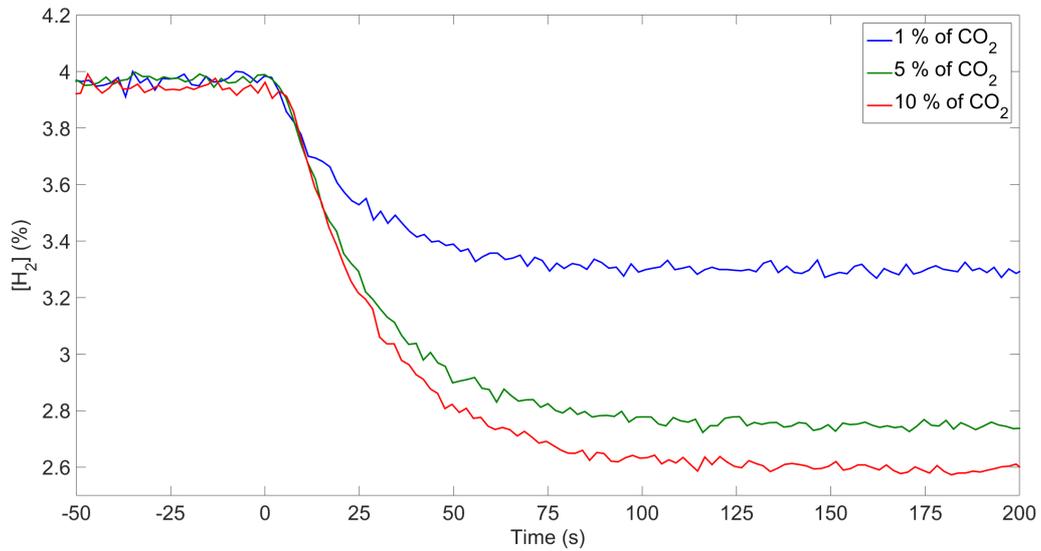

Figure 2: Evolution of the $H_2$ mixing ratio in the gaseous reactive medium with the plasma duration. Origin of the time is set as the moment when the plasma is turned on.

In (Fleury et al., 2015), the $CO_2$ consumption efficiency $e_{CO2}$ is defined according to the following equation (1):

$$e_{CO_2} = \frac{\Delta_{CO_2}}{[CO_2]_0} = \frac{[CO_2]_0 - [CO_2]_{ss}}{[CO_2]_0} \quad (1),$$

where $[CO_2]_0$ and $[CO_2]_{ss}$ represent the initial and steady-state concentrations of carbon dioxide and $\Delta_{CO2}$ represents the consumption of $CO_2$.

Similarly we defined the hydrogen consumption efficiency $e_{H2}$ with:

$$e_{H_2} = \frac{\Delta_{H_2}}{[H_2]_0} = \frac{[H_2]_0 - [H_2]_{ss}}{[H_2]_0} \quad (2),$$

where $[H_2]_0$ and $[H_2]_{ss}$ represent the initial and steady-state percentages of molecular hydrogen respectively and $\Delta_{H2}$ the consumption of $H_2$.

We defined also the elemental ratio in the fractions of gases consumed C/H:

$$\frac{C}{H} = \frac{\Delta_{CO2}}{2\Delta_{H2}} \quad (3),$$

Where $\Delta_{CO2}$ represents the absolute consumption of $CO_2$ and $\Delta_{H2}$ the absolute consumption of $H_2$.

Similarly we defined the elemental ratio in the fractions of gases consumed O/H:

$$\frac{O}{H} = \frac{\Delta_{CO2}}{\Delta_{H2}} \quad (4),$$

Where $\Delta_{CO2}$ represents the absolute consumption of $CO_2$ and $\Delta_{H2}$ the absolute consumption of $H_2$.

Table 1: Evolutions as a function of $[CO_2]_0$ of: $\Delta_{CO2}$, $e_{CO2}$, $\Delta_{H2}$, and C/H and O/H ratios in the fraction of gases consumed. The uncertainties are given as 2σ (standard deviation) and are calculated from the standard fluctuation of the mass spectrometry measurements.

| $[CO_2]_0$ (%) | $\Delta_{CO2}$ | $\Delta_{H2}$ | $e_{CO2}$ (%) | $e_{H2}$ (%) | C/H | O/H |
|---|---|---|---|---|---|---|
| 1 | 0.24 | 0.72 | 24 ± 2 | 18 ± 1 | 0.2 | 0.3 |
| 5 | 1.2 | 1.3 | 24 ± 1 | 32 ± 1 | 0.5 | 0.9 |
| 10 | 2 | 1.4 | 20 ± 1 | 36 ± 1 | 0.7 | 1.4 |

Consumption efficiencies are reported in Table 1. A first important observation is that $CO_2$ and $H_2$ are efficiently consumed whatever the experimental conditions with a relative decrease of about 20-30%. This first result is important meaning that new molecules are produced in the discharge from the reactions of these two species, involving the formation of possible organic molecules. Organic growth seems possible even in the absence of methane, with $CO_2$ as the sole carrier for organic chemistry.

The consumption efficiencies of $H_2$ and $CO_2$ are moreover rather different according to the initial $CO_2$ concentration. On the one hand the $CO_2$ relative consumption is stable at about 25%, so that when the initial $CO_2$ amount is multiplied by a factor of 10, the $CO_2$ consumed is multiplied by about the same factor. Considering the whole carbon balance among the reactants and the products in the discharge, we therefore expect a C-content similarly extended in the budget of the products. On the other hand the $H_2$ consumption strongly increases with the initial $CO_2$ concentration, but it does not compensate the larger absolute $CO_2$ consumption. This leads to an important increase, from 0.2 up to 0.7, of the total C/H ratio provided to the product budget by the reactant consumption. We therefore expect a production of organic products whose hydrogenation rate will decrease when the $CO_2$ initial concentration increases, limited by the constant 4 % $H_2$ initial concentration in the gas mixture. To study this possible change of the chemistry as a function of the initial amount of $CO_2$, we will now study the resulting products of this chemistry in a gaseous mixture made of $N_2$, $CO_2$ and $H_2$.

### 3.2. Solid phase production

A solid thin film is only observed in experiments realized with 10% of $CO_2$ on the $CaF_2$ substrates disposed on the grounded electrode. The film is organic as confirmed by its mid-IR signature presented in Figure 3.

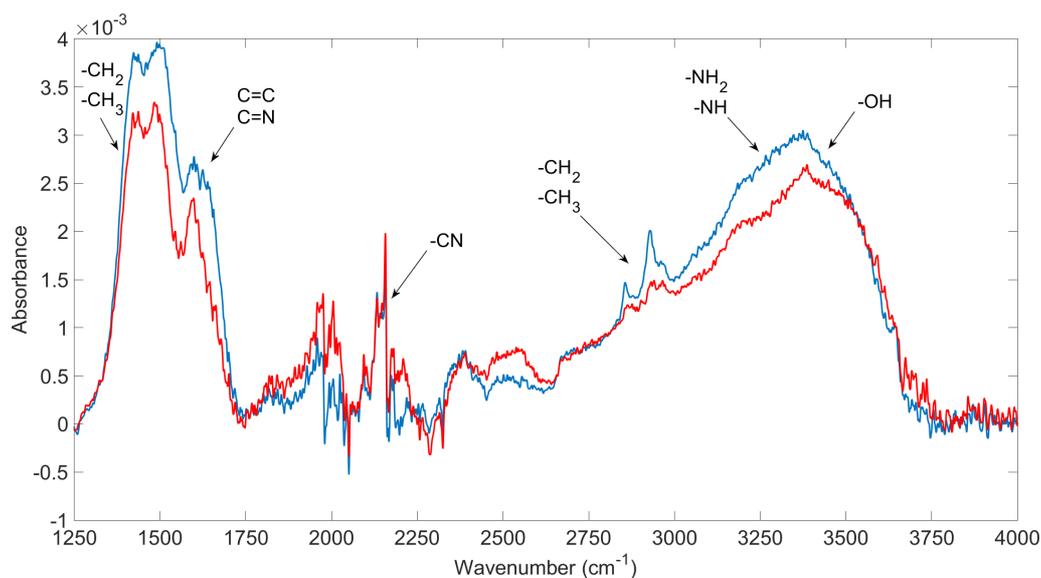

Figure 3: Infrared absorption spectra of the films deposited on two substrates during the same experiments in a $N_2/CO_2/H_2$ (86/10/4) plasma. The films were produced during 40 hours of plasma duration.

Different absorption bands characteristic of solid organics are observed in the spectrum.

First, two broad bands are observed at lower wavenumbers, centered at 1465 cm$^{-1}$ and 1630 cm$^{-1}$. The 1465 cm$^{-1}$ band can be a contribution of the asymmetric bending mode of -CH$_3$ and a scissor in plane bending mode of -CH$_2$. The second band at 1630 cm$^{-1}$ corresponds to several possible functional groups as C=N and C=C double bonds, aliphatic and aromatic -NH$_2$ or aromatic and hetero-aromatic functions.

Another small absorption band is visible at 2140 cm$^{-1}$, which corresponds to nitrile bonds -C≡N or isocyanides -N≡C.

The 2860 cm$^{-1}$ band is attributed to –CH$_2$ symmetric stretching mode. The 2930 cm$^{-1}$ and 2960 cm$^{-1}$ bands are attributed respectively to the -CH$_2$ asymmetric stretching mode and the -CH$_3$ asymmetric stretching mode.

Finally, a broad band is observed at 3380 cm$^{-1}$. This band is consistent with N-H amine bonds.

No obvious C=O carbonyl bond can be detected at ≈1700 cm$^{-1}$ in spite of the $CO_2$ gas reactant used in the experiment. The incorporation of oxygen in the solid material is not possible to confirm by mid-IR spectroscopy. There may be some hydroxyl bonds but those are ambiguously overlapping with amine functions. Nitrogen-bearing chemical functions are present involving at least nitrile or isocyanide functions. The formation of a N-rich solid highlights the possibility for an efficient ionospheric chemistry to provide prebiotic material on the early Earth.

### 3.3. Identification of the gas-phase products

We will now study the gaseous intermediate species formed during the experiments. In order to detect and identify gas species produced in low quantity, we have trapped the volatile species and accumulated them by cooling the plasma box as described in section 2.2. The plasma box is then warmed up to room temperature. A pressure increases is observed associated to the sublimation of the trapped species.

#### Ammonia: a precursor of the solid phase

The broad band observed at 3380 cm$^{-1}$ suggests an important amine content in the solid phase produced with 10% of $CO_2$. A possible precursor in the gas phase for this chemical signature could be ammonia. To confirm this, we realized an analysis of the gaseous phase by infrared spectroscopy. Figure 4 presents an infrared spectrum in the 700-1200 cm$^{-1}$ range recorded at room temperature after the release of the trapped gases.

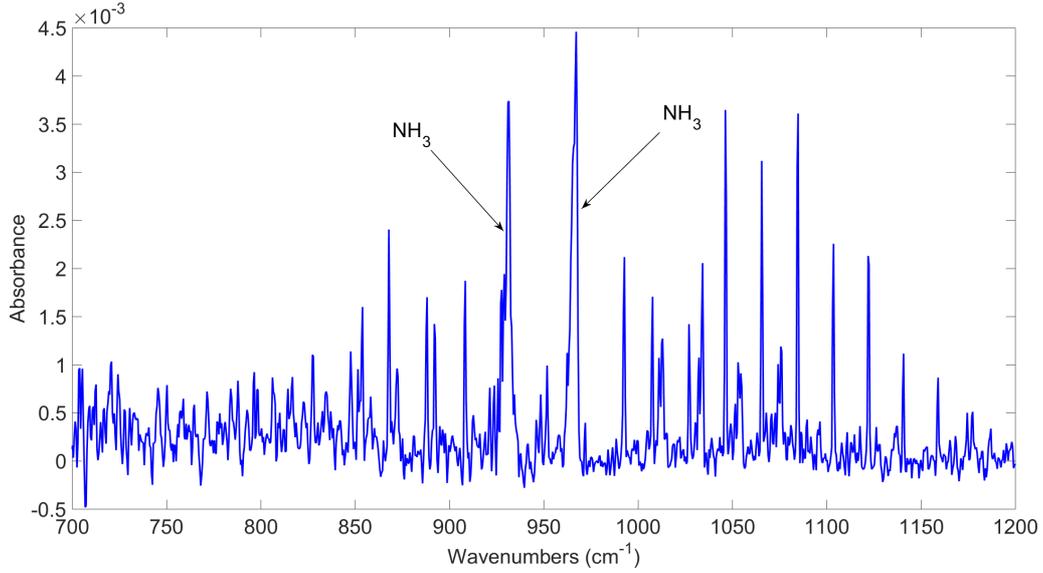

Figure 4: Infrared spectra in the 700-1200 cm$^{-1}$ range recorded at 300 K after 4 hours of cryogenic trapping in a $N_2$-$CO_2$-$H_2$ (91/5/4) plasma.

This spectrum presents two strong absorption bands at 930 cm$^{-1}$ and 967 cm$^{1}$. They corresponds to the two Q branches of the υ2 vibrational transition of $NH_3$. The existence of these two bands, characteristic of the ammonia molecule, is caused by the inversion of the molecule by quantum tunneling effect. This results in the splitting of the low energy levels in a pair with slightly different energy values. In addition, some rotational lines of the P and R branches are also observed on this spectrum. Those bands allow its identification unambiguously. Since $NH_3$ has absorption bands with no overlap over other species, it is possible to estimate its concentration using the Beer-Lambert law. The absorbance $A(\lambda)$ at a given wavelength is defined by:

$$A(\lambda) = \varepsilon(\lambda) \times l \times [C] \quad (5),$$

where $\varepsilon(\lambda)$ is the absorption cross-section of the molecule at a given wavelength, l is the path length of the beam through the gas cell and $[C]$ is the concentration of absorbing molecules in the reactor.

$$[C] = \frac{A(\lambda)}{\varepsilon(\lambda) \times l} \quad (6)$$

The ammonia absorption cross-sections have been calculated from the line by line parameters provided by the HITRAN 2012 database (Rothman et al., 2013), using the HITRAN Application Programming Interface (HAPI) (Kochanov et al., 2016). In order to overcome the difference of resolution between the database and laboratory data, we calculate the concentration from the area of the bands. For $NH_3$, integration is performed both in the 920-940 $cm^{-1}$ and in the 950-970 $cm^{-1}$ ranges.

Knowing the volume of the reactor we calculate the number of molecules of ammonia $N_{NH3}$ trapped during four hours. Table 2 presents the evolution of $N_{NH3}$ as a function of the $CO_2$ initial amount. A good agreement is found for $N_{NH3}$ calculated with the two bands.

Table 2: Evolution as a function of $[CO_2]_0$ of the number of $NH_3$ molecules formed in 4 hours of plasma duration. The uncertainties are given as 2σ (standard deviation) and are calculated from the standard fluctuations of the infrared spectroscopy measurements.

| $[CO_2]_0$ (%) | $N_{NH3}$ (molecules) 920-940 $cm^{-1}$ | $N_{NH3}$ (molecules) 950-970 $cm^{-1}$ |
|---|---|---|
| 1 | $(5.3 \pm 0.1) \times 10^{16}$ | $(6.4 \pm 0.5) \times 10^{16}$ |
| 5 | $(6.7 \pm 0.4) \times 10^{16}$ | $(6.9 \pm 0.5) \times 10^{16}$ |
| 10 | - | $(4.1 \pm 0.2) \times 10^{15}$ |

The number of ammonia molecules trapped in the 1% and 5% conditions is similar, but this number decreases by one order of magnitude for the 10% case. This result comes in front of the solid organic phase production, which is observed in the only 10% case. So compared to the lower $CO_2$ initial concentrations, in the 10% condition $NH_3$ disappears from the gas phase whereas an N-H bond signature is observed in the solid. These observations suggest that in all likelihood ammonia is a precursor for the solid phase with 10% of $CO_2$ and leads to the amine signature of the solid.

The formation of ammonia in our $CO_2$-$N_2$-$H_2$ plasma requires similar pathways as discussed in (Carrasco et al., 2012) for the case of a $N_2$-$CH_4$ plasma. Those involve NH radicals reacting with molecular hydrogen (Carrasco et al., 2012):

$$NH + H_2 \rightarrow NH_3 \quad (R1).$$

NH radicals are produced at least through three pathways considering the literature and our gaseous mixture. The first one involves radical chemistry (Carrasco et al., 2012; Mutsukura, 2001):

$$N + H \rightarrow NH \quad (R2)$$

The second one involves ion chemistry (Green and Caledonia, 1982):

$$N_2^+ + H_2 \rightarrow N_2H^+ + H \quad (R3)$$

$$N_2H^+ + e^- \rightarrow NH + N \quad (R4)$$

The last one involves radical chemistry (Dobrijevic et al., 2014) and is related to the production of water in the $CO_2$-$N_2$-$H_2$ discharge (Fleury et al., 2015):

$$N\,(^2D) + H_2O \rightarrow NH + OH \quad (R5)$$

No other molecule than water and ammonia can be detected by IR spectroscopy by lack of sensitivity. We therefore complete the gas phase analysis, first with ex-situ GC-MS and secondly by MS analysis.

*Ex situ* analysis by GC-MS of the gaseous products after cryogenic trapping

The cryogenic trapping is performed at 173 K for 8 hours of plasma duration. After the warming of the plasma box, the pressure in the reactor is about 4.1 mbar. Then the gases released are transferred into an external cold trap, resulting in a pressure of about 50 mbar in the external trap before the injection in the GC-MS. The chromatogram obtained for an initial gaseous mixture containing 5% of $CO_2$ is presented in the top of the Figure 5. All the peaks have been identified using their retention time and their mass spectra. Four species are detected.

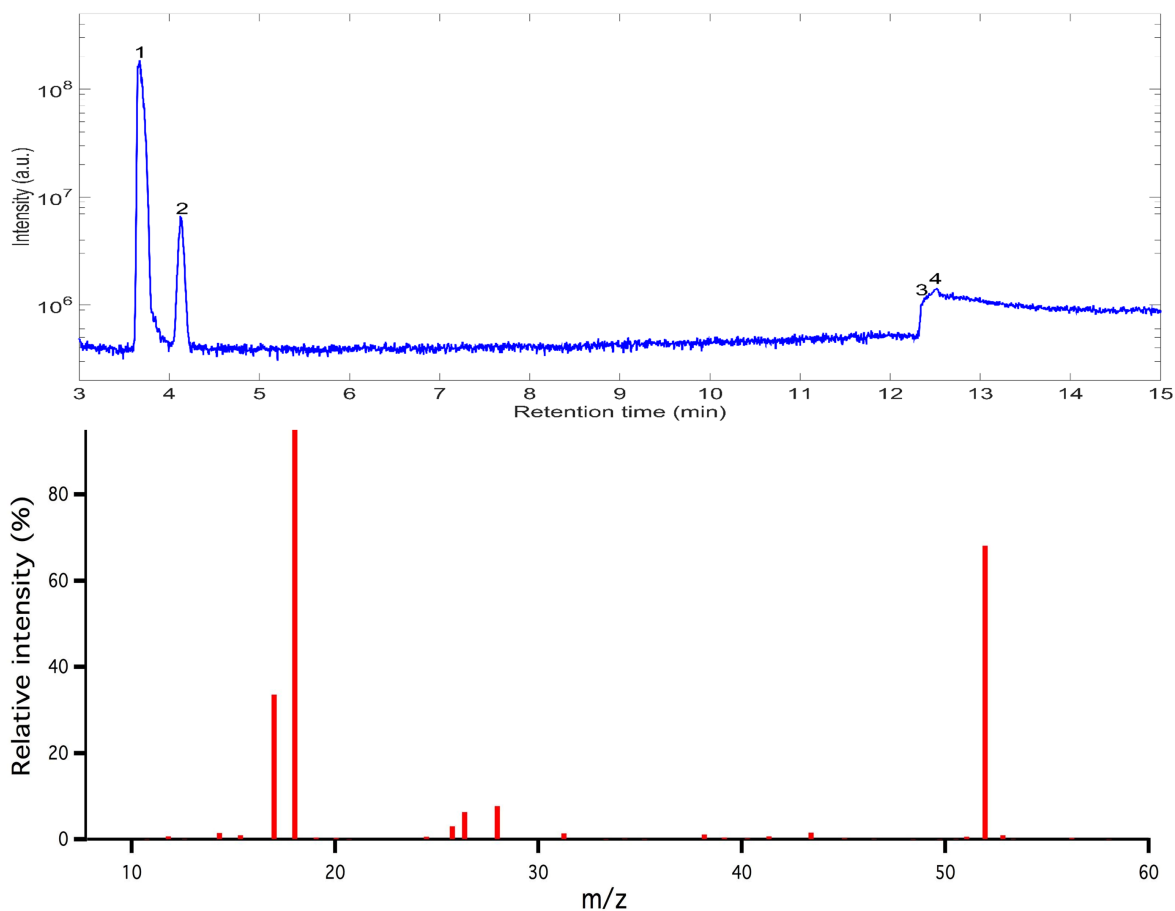

Figure 5: Top: Chromatogram of the gases trapped during an experiment with 5% of $CO_2$. Bottom: mass spectrum corresponding of the peak 4 of the chromatogram. This peak is attributed to $C_2N_2$ in agreement with the ion detected at *m/z* 52. The presence of the mass peak at *m/z* 17 and 18 is explained by the presence of water co-eluted with $C_2N_2$.

The first molecule identified is carbon dioxide, with a retention time of 3.42 min, and corresponds to a major peak in the chromatogram. This detection is surprising. Indeed, $CO_2$ is not expected to condense in these conditions of pressure and temperature. Most likely explanation, an important formation of water ice is observed during the experiment, which can trap small molecules depending on the temperature (Bar-nun et al., 1985; Notesco and Bar-Nun, 1997). This unexpected trapping of $CO_2$ has to be taken into account in the following and possibly extended to other gas products, which are not expected to condense such as hydrocarbons with 2 carbon atoms.

The second species detected is nitrous oxide $N_2O$ with a retention time of 4.06 min. $N_2O$ has previously been detected in the gaseous products in a simulation realized with the PAMPRE experiment and using an initial gaseous mixture made of $N_2$, CO and $CH_4$ (Fleury et al., 2014).

The third species detected is water with a retention time of 12.34 min. because of the saturation of the chromatographic column. This detection of water is in agreement with its previous detection (Fleury et al. 2015).

And the last detected species, co-eluted with water, is ethanedinitrile ($C_2N_2$) with a retention time of 12.52 min. The mass spectrum corresponding to this peak is presented in the bottom of the Figure **5**. The important peak at *m/z* 52, with a relative intensity of 70 %, allows an unambiguous identification of $C_2N_2$. The presence of two other important peaks at *m/z* 17 and 18 is explained by the co-elution of water with $C_2N_2$ and, which is responsible for these two peaks in the mass spectrum.

### Analysis of the trapped gaseous products by mass spectrometry

The identification of the gaseous products is completed by a direct analysis of the gas trapped released in the reactor by using mass spectrometry. Mass spectra are found similar for the 5 and the 10% $CO_2$ experiments, in agreement with the total pressure of the released products. Intensities are found lower in the 1% $CO_2$ experiment, also in agreement with the total gas pressure collected at the end of the warming. No qualitative difference is noted according to the initial $CO_2$ concentration. The further analysis is therefore focused on one condition, the 5% $CO_2$ experiment.

Figure 6 presents two mass spectra obtained with an initial $CO_2$ amount of 5%: the first spectrum is recorded at 173 K and the second after the warming of the plasma box to room temperature (294 K). As water and ammonia were previously identified and quantified by FT-IR spectroscopy, the interpretation of the spectra is achieved only for m/z > 20.

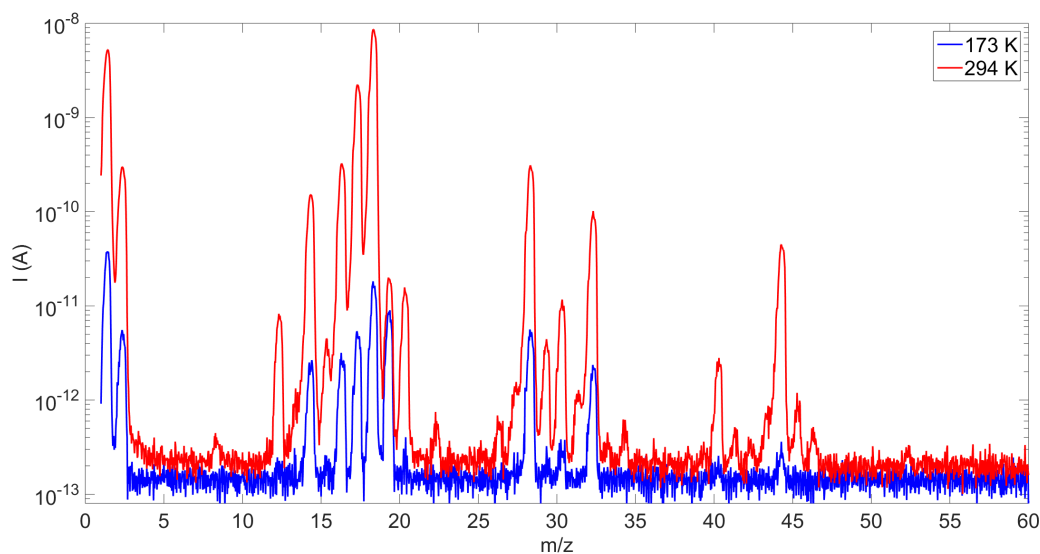

Figure 6: Mass spectra recorded at 173 K (blue) and 294 K (red) after 4 hours of gaseous trapping in $N_2$-$CO_2$-$H_2$ (91/5/4) plasma.

First, the signature of the residual air in the mass spectrometer (contributions of $N_2$, $O_2$, $CO_2$ and Ar) is identified in the mass spectrum recorded at 173 K. It corresponds to the vacuum limit of the MS pumping, at $3 \times 10^{-8}$ mbar.

At room temperature, Figure 6 shows an evolution of the mass spectrum with the one recorded at 173 K. We observe the increase of the intensity of some peaks and the apparition of new ones with masses up to 60 u. This reflects a release of gaseous species in agreement with the increase of the pressure measured in the reactor.

The low resolution of our mass spectrometer does not allow differentiating molecules with close masses. However, few molecules can be tentatively identified. As discussed previously in the GC-MS section with the detection of $CO_2$, the important formation of water ice on the walls of the plasma box impacts the species simultaneously trapped. This water trapping explains the four highest intensities for m/z > 20 observed in Figure **6** at 294 K: m/z 28, 32, 40 and 44 for $N_2$ and CO, $O_2$, Ar and $CO_2$ respectively. Indeed these molecules do not condense in the conditions of pressure and temperature of the experiment. Given the detection of $N_2O$ by GC-MS analysis, a contribution of this molecule is expected at m/z 44 in addition to the main $CO_2$ signature. Indeed, in mass spectrometry, $N_2O$ has a principal fragment at m/z 44, which is mixed with the $CO_2$ signature according to the analysis of the gaseous phase achieved by GC-MS and presented above. The second fragment is at m/z 30 and represents 31% of m/z 44 intensity. This pattern is compatible with the mass spectrum presented in Figure **6**.

A first important precursor for the formation of complex organic molecules is hydrogen cyanide HCN. Its signature in mass spectrometry is a major peak at *m/z* 27 with a relative contribution at *m/z* 26 (20%). This signature is visible on the mass spectrum in Figure 6, but is low, involving a low concentration of HCN is the gas phase. This concentration is even under the detection limit of our FTIR diagnosis so that it cannot be definitely quantified in the experiment. A pathway for HCN formation is well established in the case of the chemistry of $N_2$ and $CH_4$ in Titan's atmosphere (Hébrard et al., 2012; Krasnopolsky, 2009; Loison et al., 2015; Wilson and Atreya, 2003). It involves the key species $CH_3$ produced by the photolysis of $CH_4$ in Titan's atmosphere. This mechanism is less obvious in a $N_2/CO_2/H_2$ plasma discharge, in agreement with the low concentration of HCN observed.

At m/z 26 the ratio between the peak at 27 and 26 is larger than the ratio expected from the NIST database for HCN. This more important intensity of the *m/z* 26 peak can be explained by the contribution of all abundant nitriles with their $CN^+$ fragments even if the presence of acetylene $C_2H_2$ cannot be discarded. Indeed, if $C_2H_2$ does not condense in these conditions of pressure and temperature but could be trapped in water ice.

Given the trapping of $N_2$ or CO revealed at *m/z* 28, the peak at *m/z* 29 could be mainly attributed to the isotopologues $^{14}N^{15}N$ or $^{13}C^{16}O$. In our experiment, the intensity ratios $I_{29}:I_{28}$ are 100:1.13, while the expected $I_{29}:I_{28}$ ratios for natural abundance isotopologues of $N_2$ and CO are 100:0.75 and 100:1.12 respectively as given by the NIST database. The ratio measured in our experiment is in good agreement with the one expected for CO, which is so the most likely candidate contributing to these peaks. The presence of CO in the mass spectrum can result of the dissociation of $CO_2$ in the mass spectrometer or of its trapping during the experiment, CO being a major product of the $CO_2$ dissociation in the plasma.

The peak at *m/z* 30 is important in the mass spectrum observed at 294 K. There are two possible species at this mass: formaldehyde $H_2CO$ and ethane $C_2H_6$ with a similar explanation for ethane trapping as for acetylene. Given the abundance of the reactants $CO_2$ and $H_2$ in the gas mixture, formaldehyde would be expected as an important contributor. $H_2CO$ are at 29 and 30 amu with a 29:30 ratio of 100:58. This is in disagreement with the ratio observed in the mass spectrum of the Figure 6, where the main contribution to the peak at 29 amu is attributed to $^{13}CO$. Our results point out a possible production of ethane even in the oxidant conditions used in the present work.

At *m/z* 41, one possible species is acetonitrile $CH_3CN$, in agreement with the formation of HCN, which is a precursor of $CH_3CN$ (Dobrijevic and Dutour, 2006; Gautier et al., 2011).

The intensities of the peaks at *m/z* 45 and 46 are consistent with isotopes of $CO_2$.

And a last peak is observed at *m/z* 52 in agreement with the detection of ethanedinitrile $C_2N_2$ by GC-MS.

In conclusion in situ mass spectrometry completes our overview of the gas products by pointing out nitrile molecules. Among them hydrogen cyanide HCN is an important molecules for prebiotic chemistry, for example involved with ammonia $NH_3$ in the Strecker amino acid synthesis (Strecker, 1854).

## 4. Discussion

The possible formation of organic aerosols in the atmosphere of the early Earth is of the prime interest for the comprehension of the environment of the Earth during the Hadean and the Archean eons. First of all, the formation of fractal organic hazes can provide an ultraviolet shield for the early Earth (Wolf and Toon, 2010) against the high UV flux of the young Sun (Claire et al., 2012). This shield affects the composition of the atmosphere protecting some constituents of the atmosphere such as ammonia from photochemical destruction (Sagan and Chyba, 1997). Furthermore, the formation of haze would affect the climate of the early Earth providing an antigreenhouse effect (Haqq-Misra et al., 2008; Hasenkopf et al., 2011; McKay et al., 1999).

In our gaseous mixture, solid organic products are only observed as deposited on a substrate. The solid formation seems to be promoted on the substrate with the formation of an organic thin film, which has grown on a substrate placed on the grounded electrode of the plasma device. On the contrary, the formation of spherical shaped individual grains is not observed. The mechanism of formation of these two types of solid in the plasma is not fully known. Both grow from reactive gas species present in the plasma. For the formation of grains in the plasma, the process starts with the formation of nanometer-size monomers, which aggregate to form larger particles with a negative charge when the density of the monomers in the plasma reaches a critical value. Then the particles grow by deposition of species present in the gaseous phase (Wattieaux et al., 2015). The hypothesis, which can be done to explain the absence of these grains in our gaseous mixture compared to $CH_4$-rich experiments (Szopa et al. 2006) is that the lower reactivity of $CO_2$ does not allow to reach the critical value of monomers in the plasma necessary to start the growth of larger particles.

With such a mechanism, we expect that the water formation observed in this gaseous mixture would sustain the formation of solid organic in the case of the atmosphere of the early Earth. Indeed, as discussed in Fleury et al. 2015, water formed above the troposphere leads to the formation of high altitude clouds analogues to the present PSC and PMC. These clouds are composed of water ice, which would play the role of nucleus cores, promoting a heterogeneous nucleation of organic molecules at the surface of ice grains.

Moreover, the formation of organic volatiles in interaction with water ice can also be important in the perspective of a potential enrichment of the lower atmospheric layers of the early Earth by prebiotic molecules formed in the upper atmosphere. Indeed, these molecules could be trapped in the water ice clouds and transport to the lower atmospheric layers involving an important source if nitrogenized molecules for the lower atmospheric layers.

## Conclusion

We have shown that an organic growth is possible in an oxidized atmosphere made of $N_2$, $CO_2$ and $H_2$ simulating the early Earth atmosphere. This suggests that an organic growth can been sustained only by carbon composing $CO_2$ in the primitive atmosphere of the Earth without methane. Firstly this results in the formation of volatiles molecules with the identification of $H_2O$, $NH_3$, $C_2N_2$, $N_2O$ and the possible detection of HCN. The detections of these molecules are interesting in a prebiotic chemistry point of view. Indeed, they are reactive molecules, which conduct to the formation of more complex molecules and HCN and $NH_3$ are involved in the formation of amino acids and nitrogenous bases. Secondly we have shown that in the absence of methane $CO_2$ also plays a role in the formation of organic compounds. Indeed, if the composition of the gaseous phase does not change as a function of the $CO_2$ initial amount, an increase of the $CO_2$ content promotes the formation of organic compounds highlighting that a high level of $CO_2$ in the primitive atmosphere of the Earth can promote organic chemistry in such an atmosphere. And finally, an organic thin film formation is observed in the experiments requiring a threshold concentration of $CO_2$. Ammonia has also been shown to be precursor for this solid formation. This highlights the important complexity of the organic chemistry, which could be initiated in the early Earth atmosphere from $N_2$, $CO_2$ and $H_2$.

Another result, from this work is that in our gaseous mixture, solid organic products are only observed as thin films deposited on a substrate and a low quantity. In such mechanism, the water formation observed in our experiment can play a role in the formation of solid organic in the atmosphere of the early Earth. Indeed, water formed above the troposphere would conduct to the formation of high altitude clouds analogues to the present PSC and PMC. These clouds could then play the role on nucleus cores, promoting a heterogeneous nucleation of organic at the surface of ice grains.

In this work we have investigated the influence of the $CO_2$ concentration on the chemistry but the hydrogen concentration in the primitive atmosphere of the Earth is also debated. Further work will be necessary to understand the influence of the hydrogen concentration of the early Earth's atmospheric reactivity for different concentrations of carbon dioxide.

# Acknowledgments

B. Fleury acknowledges the Ile-de-France region (DIM ACAV) for his thesis funding. N. Carrasco acknowledges the support of the European Research Council (ERC Starting Grant PRIMCHEM, grant agreement n°636829).